\documentclass{revtex4}
\usepackage{epstopdf,amsmath}
\usepackage{graphicx}

\newcommand{\Array}[2]{\left(\begin{array}{#1}#2\end{array}\right)}

\begin{document}

\title{Dual Implications of Quark Mass Hierarchies to Flavor Structure}     
\author{Ying Zhang$^{1,2}$\footnote{Corresponding author. E-mail:hepzhy@mail.xjtu.edu.cn.}}
\address{$^1$School of Science, Xi'an Jiaotong University, Xi'an, 710049, China\\
		$^2$ Institute of Modern Physics, Xi'an Jiaotong University, Xi'an, 710049, China}

\begin{abstract}
To solve the mystery of flavor structure, we demonstrate two implications emerging from the hierarchical masses of quarks: one for the mass matrix itself and one for the CKM mixing. These implications naturally lead to a non-redundant, ordered, and family-unified quark flavor structure, which serves as a candidate to replace the unclear Yukawa interactions of the Standard Model.
\\
\textbf{Keywords:} flavor structure; sub-unitarity; the flat matrix; mass hierarchy
\end{abstract}

	\maketitle
\section{Motivation}
Although the Standard Model (SM) has been extensively validated by high-energy phenomenology, many fundamental questions regarding its flavor structure remain unanswered \cite{FroggattNPB1979, FeruglioEPJC2015, XingPR2020, KingRPP2013}. It is unknown why the quark CKM mixing angles are small while the lepton PMNS mixing exhibits two large mixing angles, and whether some underlying relations exist between quark masses and their mixing in the weak charged current.
In the SM, these flavor problems originate from the Yukawa interaction, whose complex couplings determine quark masses and the CKM matrix. Flavor phenomenology is ultimately calculated in terms of quark masses and CKM parameters rather than the Yukawa couplings themselves. 
No theoretical guidance for these couplings, neither their values nor their possible structure, is provided by the SM. 
Deciphering the flavor structure has thus become a central mission for completing the Yukawa interaction and understanding the origin of mass hierarchy and flavor mixing.

To construct a desired flavor structure, the first step is to establish a clear description of flavor phenomenology that is free from redundant parameters. This is crucial for two reasons. First, complex, family-dependent Yukawa couplings introduce numerous unobservable parameters, which obscure the underlying clues to fermion masses and mixing. Second, the absence of definitive evidence for new physics beyond the SM to date strongly suggests that introducing new interactions or particles to explain flavor questions is unnecessary.

Historically, texture zero matrices \cite{FritzschSS2017, FritzschPPNP2000} have provided an exploratory framework for relating CKM mixing angles to quark mass ratios, achieved by reducing quark mass matrices to a minimal form characterized by vanishing entries. Nevertheless, this ansatz faces increasing tension with precision CKM measurements, and its inherent basis dependence calls into question the physical significance of its postulated zeros.
A recent study has proposed a minimal flavor structure based on hierarchical masses, achieving a highly precise description of quark masses and CKM mixing with non-redundant parameters \cite{ZhangEPL2024}. 

In this paper, by extending the implication of mass hierarchy in mass pattern to the CKM mixing, we focus on the dual role of mass hierarchy. The concept of sub-unitarity is proposed as a hidden self-consistency characteristic of the CKM matrix for the first two quark families, It provids a benchmark for evaluating flavor models. This framework illuminates the origin of the small CKM mixing angles and offers insights into the contrasting pattern of large PMNS mixing angles. While this paper focuses primarily on the quark sector, we briefly discuss leptons due to remaining uncertainties in the nature and masses of neutrinos.
This paper is organized as follows. In Sec. \ref{sec.RevMassPattern}, we logically derive the factorized form of the quark mass matrix as the first implication from hierarchical masses by systematically recasting the key arguments in the minimal flavor structure. 
In Sec. \ref{sec.implicationMixing}, the concept of sub-unitarity as the second implication is introduced, demonstrating its emergence as an effective theory below the third-family mass scale. We combine both implications to construct a unified flavor structure in Sec. \ref{sec.FlatPattern}, culminating in the flat matrix as the most natural pattern. Sec. \ref{sec.HighOrderCorrection} addresses hierarchy corrections beyond the leading order. Sec. \ref{sec.fit} presents phenomenological fits, and Sec. \ref{sec.sum} summarizes our conclusions.

\section{First Implication: The Factorized Mass Matrix}
\label{sec.RevMassPattern}
\subsection{Redundant Right-Handed Transformations}
\label{sec.generallFramework}
In the SM, the quark mass matrix arises from complex Yukawa couplings after electroweak symmetry breaking. For an arbitrary complex mass matrix $M^q$ (where $q=u,d$ denotes up- and down-type quarks respectively), diagonalization is achieved through a bi-unitary transformation:
\begin{eqnarray}
U_L^qM^q(U_R^q)^\dag=\textrm{diag}(m_1^q,m_2^q,m_3^q)
\end{eqnarray}
 The left-handed and right-handed unitary matrices $U_L^q$ and $U_R^q$ are determined by diagonalizing the Hermitian combinations $M^q(M^q)^\dag$ and $(M^q)^\dag M^q$ respectively
\begin{eqnarray}
&& U_L \Big[M^q(M^q)^\dag\Big] (U_L^q)^\dag=\textrm{diag}\Big((m_1^q)^2,(m_2^q)^2,(m_3^q)^2\Big)
\label{eq.ULdef}
\\
&& U_R \Big[(M^q)^\dag M^q\Big] (U_R^q)^\dag=\textrm{diag}\Big((m_1^q)^2,(m_2^q)^2,(m_3^q)^2\Big)
\end{eqnarray}

In flavor phenomenology, all physical observables are six quark masses, three CKM mixing angles, and one CP-violating phase. The right-handed rotations 
$U_R^q$ are non-physical, as they do not contribute to these observables, i.e.,
\begin{itemize}
	\item[1.] In the charged weak current, when expressing gauge fields in the mass basis, only the left-handed rotations appear
$V_{CKM}=U_L^u(U_L^d)^\dag$;
	\item[2.] For an arbitrary unitary matrix $U'$, the mass matrices $M^q$ and $M^qU'$ yield identical physical masses and left-handed rotations. This freedom allows us to fix the unphysical degrees of freedom in $U_R^q$.  
\end{itemize}

Without loss of generality, we adopt the convention $U_R^q=U_L^q$ throughout this paper.
For a non-hermitian mass matrix $\tilde{M}^q$ appearing in the literature, it is equivalent to a Hermitian matrix $M^q$ defined by 
\begin{eqnarray}
	M^q=(U_L^q)^\dag \textrm{diag}(m_1^q,m_2^q,m_3^q)U_L^q
\end{eqnarray}
where $U_L^q$ is obtainded from diagonalizing $\tilde{M}^q(\tilde{M}^q)^\dag$ as in Eq. (\ref{eq.ULdef}). 

\subsection{Hierarchical Mass Structure}
Quark masses have been measured experimentally (see Tab. \ref{tab.quarkmass}).
\begin{table}[htp]
\caption{Quark Masses. $m_1^u,m_1^d$, and $m_2^d$ (up, down and strange quark masses) are $\overline{\textrm{MS}}$ masses at scale $\mu=2$ GeV. $m_2^u$ and $m_3^d$ (charm and bottom) are $\overline{\textrm{MS}}$ masses renormalized at their own masses. The $m_3^u$ (top) comes from direct measurements \cite{PDG2024}.}
\begin{center}
\begin{tabular}{c|c|c|c}
\hline\hline
Family & 1st & 2nd & 3rd
\\
\hline
up-type &  $m_1^u=2.16\pm0.07$ MeV & $m_2^u=1.2730\pm0.0046$ GeV & $m_3^u=172.56\pm0.31$ GeV
\\
\hline
down-type &  $m_1^d=4.70\pm0.07$ MeV & $m_2^d=93.5\pm0.8$ MeV & $m_3^d=4.183\pm0.007$ GeV
\\
\hline \hline
\end{tabular}
\end{center}
\label{tab.quarkmass}
\end{table}%
Within each quark type, the family masses exhibit a striking hierarchical pattern:
\begin{eqnarray}
	m_1^u\ll m_2^u \ll m_3^u,~~~~
	m_1^d\ll m_2^d \ll m_3^d.
\end{eqnarray}
These relations describe family connections within the same type of quarks, not involving different types of quarks.
They provide a crucial clue for decoding the quark mass matrices $M^q$.
The hierarchical relations can be quantified by defining mass ratios within each quark type:
\begin{eqnarray}
	h_{12}^q=\frac{m^q_1}{m^q_2},~~
	h_{23}^q=\frac{m^q_2}{m^q_3}.
\end{eqnarray}
In the mass hierarchy limit, we have $h_{23}^q,h_{12}^q\rightarrow 0$.

The mass matrix $M^q$ can be reconstructed in terms of the hierarchy parameters $h_{ij}^q$. 
Focusing on the hierarchy, we normalize $M^q$ by the total family mass $\sum_{i=1,2,3} m^q_i$, yielding the diagonal eigenvalues:
\begin{eqnarray}
\frac{1}{\sum_i m^q_i}M^q\sim\Array{ccc}{h_{12}^qh_{23}^q && \\ &h_{23}^q& \\ && 1-h_{23}^q-h_{12}^qh_{23}^q}
\end{eqnarray}
Consequently, the normalized mass matrix admits an expansion in powers of $h_{12}^q$ and $h_{23}^q$
\begin{eqnarray}
\frac{1}{\sum_i m^q_i}M^q=M^q_0+h_{23}^qM^q_1+h_{12}^q h_{23}^q M^q_{21}+(h_{23}^q)^2 M^q_{22}+\mathcal{O}(h^3).
\end{eqnarray}
Here, $M_0^q$ is the normalized leading-order mass matrix, $M_1^q$ is 1-order correction, and $M_{2i}^q$ are 2-order corrections.
\subsection{Leading-Order Structure}
We now investigate the leading-order term $M^q_0$, namely the normalized quark mass matrix in the hierarchy limit $h_{23}^q\rightarrow 0$.
Since 
\begin{eqnarray}
	\lim_{h_{23}^q\rightarrow0}\textrm{diag}\Big(h_{12}^qh_{23}^q,~~h_{23}^q,~~1-h_{23}^q-h_{12}^qh_{23}^q\Big)=
	\textrm{diag}\Big(0,~~0,~~1\Big)
\end{eqnarray}
$M_0^q$ can be reconstructed as:
\begin{eqnarray}
M_0^q= (U_L^q)^\dag\Array{ccc}{0 && \\ & 0 & \\ && 1}U_L^q.
\end{eqnarray}
The eigenvalues $(0,0,1)$ suppress many elements of $U_L^q$. Only the third-row elements $U_{L,3i}^q$ contribut to $M_0^q$.
In general, these elements can be parameterized as:
\begin{eqnarray}
\frac{U_{L,31}^q}{U_{L,33}^q}=l_1e^{i\eta^q_1},~~
\frac{U_{L,32}^q}{U_{L,33}^q}=l_2e^{i\eta^q_2},~~
U_{L,33}^q=l_0e^{i\eta^q_0},
\end{eqnarray}
with three real phased $\eta_i^q$ and three real modulus $l_i$.
Thus, $M_0^q$ can be expressed in a factorized form:
\begin{eqnarray}
M_0^q=(K_L^q)^\dag M_N^q K_L^q.
\label{eq.factorizedM}
\end{eqnarray}
Here, $K_L^q$ is a diagonal phase matrix: 
\begin{eqnarray}
K_L^q=\textrm{diag}\Big(e^{i\eta^q_1},e^{i\eta^q_2},1\Big).
\label{eq.KLdef}
\end{eqnarray}
Note that $\eta_0^q$ plays no role here.
The complex phases in $K_L^q$ provide the origin of CP violation, which will be discussed in the next subsection.

The matrix $M_N^q$ in Eq. (\ref{eq.factorizedM}) is a real symmetric matrix
\begin{eqnarray}
	M_N^q=\frac{1}{l_1^2+l_2^2+1}\Array{ccc}{l_1^2 & l_1l_2 & l_1 \\
	l_1l_2 & l_2^2 & l_2 \\
	l_1 & l_2 & 1}
\label{eq.MnDef}
\end{eqnarray} 
Note that $l_0$ satisfies the unitarity condition
\begin{eqnarray}
	l_0=\frac{1}{\sqrt{l_1^2+l_2^q+1}}.
\end{eqnarray}
Thus, the real $l_1$ and $l_2$ control the pattern of the mass matrix and determine the quark eigenvalues.  So, we refer to $M_N^q$ as the pattern matrix.
Its elements satisfy the following relations:
\begin{eqnarray}
	&&\frac{M_{N,11}^q}{M_{N,31}^q}=\frac{M_{N,12}^q}{M_{N,32}^q}=\frac{M_{N,13}^q}{M_{N,33}^q},
	\label{eq.MNrelation1} \\
	&&\frac{M_{N,21}^q}{M_{N,31}^q}=\frac{M_{N,22}^q}{M_{N,32}^q}=\frac{M_{N,23}^q}{M_{N,33}^q}.
	\label{eq.MNrelation2} 
\end{eqnarray}

The factorized mass matrix in Eq. (\ref{eq.factorizedM}) constitutes the first implication from hierarchical quark masses. It emerges in the limit of $h_{23}^q\rightarrow 0$, independently of any assumption about $h_{12}^q$. (This allows the pattern to also apply to normal ordering Dirac neutrinos.)

\subsection{Implication: The Yukawa Basis}
\label{sec.YukawaBasis}
The isolation of complex phases in $K_L^q$ offers a profound insight: beyond providing the origin of CP violation in the CKM matrix, it suggests a new perspective on the Yukawa interaction.
In the SM, quark fields are initially expressed as gauge eigenstates of $SU(3)_c\times SU(2)_L\times U(1)_Y$. 
In addition to gauge interactions, the SM also includes Yukawa interactions between chiral fermions and the Higgs boson. 
Unlike gauge couplings, Yukawa couplings are not determined by the gauge principle. When describing the Yukawa term in the gauge basis, Yukawa couplings appear as out-of-order, family-dependent complex numbers.

The factorized form in Eq. (\ref{eq.factorizedM}) suggests the existence of a more natural basis, the Yukawa basis, in which the Yukawa couplings assume a clear and simple structure.  Defining quark fields in this new Yukawa basis, labeled by superscript ${(Y)}$, as
\begin{eqnarray}
q_i^{(Y)}=(K_L)_{ij}q_j,
\end{eqnarray}
after electroweak symmetry breaking, the Yukawa mass terms become
\begin{eqnarray}
\mathcal{L}_Y=-y^uM_N^u\bar{q}^{(Y)}_L\tilde{H}u^{(Y)}_R-y^dM_N^d\bar{q}^{(Y)}_L{H}d_R^{(Y)}.
\end{eqnarray}
Here, the couplings $y^q$ are family-independent and control the total family mass
\begin{eqnarray}
	\sum_im_i^q=y^q\frac{v_0}{\sqrt{2}}.
\end{eqnarray}
Now, in the hierarchy limit, we obtain Yukawa terms with real couplings exhibiting an order structure governed by $M_N^q$.

\section{Second Iimplicationmplication: Sub-Unitarity of Quark Mixing}
\label{sec.implicationMixing}
\subsection{Unitarity of the CKM Matrix}
In the SM, the quark mixing appears in the charged weak current when quarks are transformed from the weak gauge basis to the mass basis. The CKM mixing matrix is
\begin{eqnarray}
V_{CKM}=U_L^u(U_L^d)^\dag.
\end{eqnarray}
Theoretically, the unitarity of $V_{CKM}$ is guaranteed by the unitary transformations $U_L^q$. 
Experimentally, precision tests confirm unitarity to high accuracy within $\leq 0.1\%$ \cite{PDG2024}:
\begin{eqnarray}
	&&\sum_i|V_{CKM,1i}|^2=0.9984\pm0.0007~(\textrm{1st row}),
	\\
	&&\sum_i|V_{CKM,2i}|^2=1.001\pm0.012~(\textrm{2nd row}),
	\\
	&&\sum_i|V_{CKM,3i}|^2=0.9971\pm0.0020~(\textrm{3rd row}),
	\\
	&&\sum_i|V_{CKM,i2}|^2=1.003\pm0.012~(\textrm{2nd column}).
\end{eqnarray}
Current constraints from electroweak precision data, rare decays, and direct searches strongly limit any deviations. It implies that the three quark families are complete and no significant mixing with additional families exists.

\subsection{The Sub-Unitarity Structure}
\begin{figure}[htbp]
\begin{center}
\includegraphics[scale=0.25]{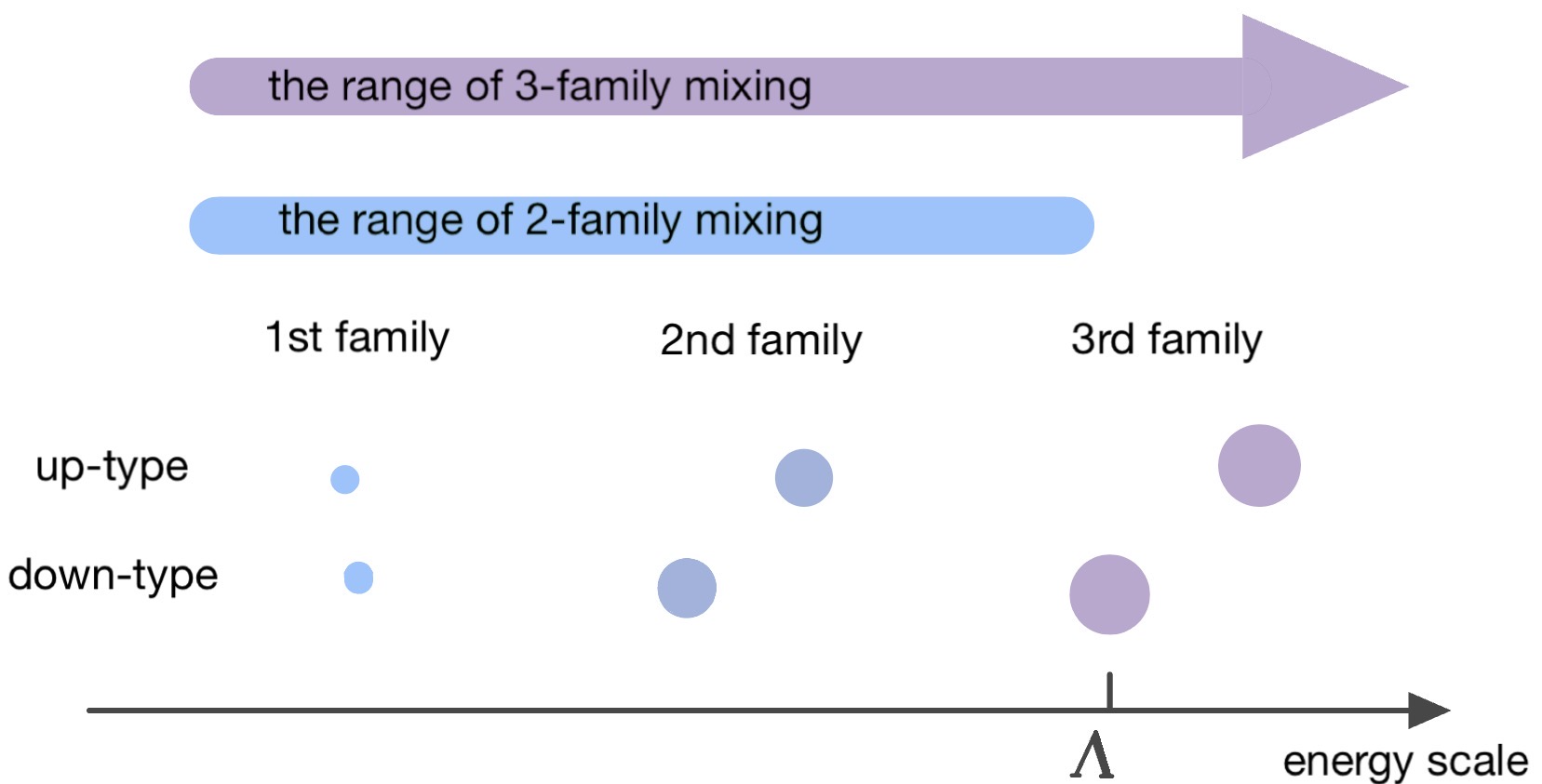}  
\caption{Quark Mass Scale}
\label{fig.quarkmass}
\end{center}
\end{figure}

Now consider $V_{CKM}$ in the mass hierarchy.
Because of $m_1^u,m_1^d, m_2^u,m_2^d\ll m_3^u,m_3^d$, there is a large gap between the third family and the first two families (see Fig. \ref{fig.quarkmass}).
Below the energy scale $\Lambda\sim m_3^d$ (bottom quark mass), the weak interaction of quarks can be described by an effective Lagrangian $\mathcal{L}_{EW}^{(2F)}$ for the first two families.  The $\mathcal{L}_{EW}^{(2F)}$ can be obtained by integrating out the third family from the SM Lagrangian $\mathcal{L}_{EW}$:
\begin{eqnarray}
\int\Pi_{i=1,2}\mathcal{D}q_{i,L}^u\mathcal{D}q_{i,L}^dq_{i,R}^u\mathcal{D}q_{i,R}^d e^{\int{d^4x}\mathcal{L}_{EW}^{(2F)}}
=\int \Pi_{i=1,2,3}\mathcal{D}q_{i,L}^u\mathcal{D}q_{i,L}^dq_{i,R}^u\mathcal{D}q_{i,R}^de^{\int{d^4x}\mathcal{L}_{EW}}
\end{eqnarray}
The leading correction to the 2-family CKM matrix from the third family appears only at the loop level. Consequently, the CKM matrix exhibits approximate unitarity for the first two families. This property is called sub-unitarity of the CKM matrix.

At the tree level, $2\times 2$ mixing matrix $V_{CKM}^{(2F)}$ can approximately expressed in terms of 3-family $V_{CKM}$ as
\begin{eqnarray}
V^{(2F)}_{CKM,ij}=V_{CKM,ij},~~\textrm{for}~i,j=1,2.
\label{eq.v2fdef}
\end{eqnarray}
The sub-unitarity requires that $V_{CKM}^{(2F)}$ tends to a unitary matrix in the limit of $m_3^u,m_3^d\gg m_2^u,m_2^d$:
\begin{eqnarray}
\lim_{{m_3^u,m_3^d\gg m_2^u,m_2^d}}V_{CKM}^{(2F)}(V_{CKM}^{(2F)})^\dag=\textbf{1}_{2\times 2}.
\end{eqnarray}

The above analysis is supported by experiment. 
Using Eq. (\ref{eq.v2fdef}), $V_{CKM}^{(2F)}$ is evaluated in the standard parameterizations
\begin{eqnarray}
V^{(2F)}_{CKM,11}&=&c_{12}c_{13},
\\
V^{(2F)}_{CKM,12}&=&s_{12}c_{13},
\\
V^{(2F)}_{CKM,21}&=&-s_{12}c_{23}-c_{12}s_{23}s_{13}e^{i\delta_{CP}},
\\
V^{(2F)}_{CKM,22}&=&c_{12}c_{23}-s_{12}s_{23}s_{13}e^{i\delta_{CP}},
\end{eqnarray}
where $s_{ij}=\sin\theta_{ij}$, $c_{ij}=\cos\theta_{ij}$, and $\delta_{CP}$ is the CP-violating phase.
The current data \cite{PDG2024} yields
\begin{eqnarray}
V_{CKM}^{(2F)}(V_{CKM}^{(2F)})^\dag=\Array{cc}{1.000& -(0.6420-1.423i)\times 10^{-4} \\ 
	- (0.6420+1.423i)\times 10^{-4}  & 0.9983}.
\end{eqnarray}
demonstrating excellent approximate sub-unitarity.

\subsection{Implications for Mixing Angles}
The unitarity of $V_{CKM}^{(2F)}$ has important implications for the origin of mixing angles.
In the two-family space spanned by the first two quark generations, $V_{CKM}^{(2F)}$ as a general $2\times 2$ unitary matrix contains 4 real parameters. After rephasing the four quark fields (which eliminates three phases, leaving one global phase irrelevant), the only remaining physical parameter is just a single mixing angle $\theta_{12}$. Thus, in the hierarchy limit, $V_{CKM}^{(2F)}$ reduces to a real orthogonal rotation
\begin{eqnarray}
\lim_{{m_3^u,m_3^d\gg m_2^u,m_2^d}}V_{CKM}^{(2F)}=\Array{cc}{c_{12} & s_{12} \\ -s_{12} & c_{12}}.
\label{eq.2Fmixing0}
\end{eqnarray}
This suggests that the full $3\times 3$ CKM matrix can be expanded as a series in the small hierarchy parameters $h_{ij}^q$:
\begin{eqnarray}
		V_{CKM}=\Array{ccc}{c & s & 0 \\ -s & c & 0 \\ 0 & 0 & 1}+h_{23}^uV_{1}^u+h_{23}^dV_{1}^d+\mathcal{O}(h^{2})
\end{eqnarray}
where $V_1^{u}$ and $V_1^{d}$ are first-order correction matrices.
Analysis of the magnitudes of CKM elements reveals
\begin{eqnarray}
		Re[V_{CKM,ij}]&\sim& \mathcal{O}(h^0),~~\textrm{for~}i,j=1,2 \textrm{ and } i=j=3,
		\\
		Re[V_{CKM,ij}]&\sim& \mathcal{O}(h^1),~~\textrm{for~other~combinations},
		\\
		Im[V_{CKM,ij}]&\sim& \mathcal{O}(h^1),~~\textrm{for~all~}i,j.
\end{eqnarray}
From these, we obtain the behavior of the standard CKM mixing angles
\begin{eqnarray}
		s_{12}^2&=& \frac{|V_{CKM,12}|^2}{1-|V_{CKM,13}|^2} \sim \mathcal{O}(h^0),
		\\
		s_{23}^2&=& \frac{|V_{CKM,23}|^2}{1-|V_{CKM,13}|^2} \sim  \mathcal{O}(h^2),
		\\
		s_{13}^2&=&|V_{CKM,13}|^2 \sim \mathcal{O}(h^2).
\end{eqnarray}
This constitutes a crucial result: the mixing angles $\theta_{13}$ and $\theta_{23}$ vanish in the hierarchy limit $h_{23}^q\rightarrow 0$.
Their small observed values arise directly from corrections due to the mass hierarchy:
\begin{eqnarray}
	\theta_{13},\theta_{23}\sim \mathcal{O}(h_{23}^q).
\end{eqnarray}

For the CP-violating phase, the Jarlskog invariant exhibits the parametric behavior:
\begin{eqnarray}
		J_{CP}
		=s_{13}c_{13}^2s_{23}c_{23}s_{12}c_{12}s_{\delta}
		\sim\mathcal{O}(h^2).
\end{eqnarray}
It implies that the CP-violating phase itself is of order unity
\begin{eqnarray}
		\delta_{CP}\sim \mathcal{O}(h^0).
\end{eqnarray}

These insights represent the second implication arising from hierarchical quark masses: any viable quark flavor model must not only achieve a precise fit to quark masses and mixing but also accurately explain the parametric relations linking mixing angles to the mass hierarchy.
    

\section{A Unified Flavor Structure}
\label{sec.FlatPattern}
The two implications, the factorized mass matrix and the sub-unitarity of CKM mixing, must be combined into a coherent framework. In this section, we construct such a unified flavor structure.

\subsection{Family Symmetry and Its Breaking}
The pattern matrix $M_N^q$ in Eq. (\ref{eq.factorizedM}) exhibits an important symmetry. 
For arbitrary parameters $l_1$ and $l_2$, $M_N^q$ is invariant under  an $SO(2)^q$ rotation \cite{ZhangEPJC2025}
\begin{eqnarray}
R_n(\theta)M_N^qR_n^T(\theta)=M_N^q
\label{eq.RMNR}
\end{eqnarray}
where $R_n(\theta)$ is
\begin{eqnarray}
R_n(\theta)=\Array{ccc}{ n_x^2(1-c_\theta)+c_\theta & n_xn_y(1-c_\theta)+n_zs_\theta & n_xn_z(1-c_\theta)-n_ys_\theta \\
	n_xn_y(1-c_\theta)-n_zs_\theta & n_y^2(1-c_\theta)+c_\theta & n_yn_z(1-c_\theta)+n_xs_\theta \\
	n_xn_z(1-c_\theta)+n_ys_\theta & n_yn_z(1-c_\theta)-n_xs_\theta & n_z^2(1-c_\theta)+c_\theta}.
\end{eqnarray}
along the rotation axis $\textbf{n}=(n_x,n_y,n_z)=\frac{1}{\sqrt{l_1^2+l_2^2+1}}(l_1,l_2,1)$.
Since the $SO(2)^q$  symmetry acts only on the quark mass matrix, it is called a family symmetry to differentiate  from flavor symmetry existing in the full electroweak Lagrangian.

 The diagonalization of $M_N^q$ can be achieved by an orthogonal transformation $S^q$
\begin{eqnarray*}
	S^qM_N^q(\theta)(S^q)^T=\textrm{diag}(0,0,1).
\end{eqnarray*}
with
\begin{eqnarray}
	S^q=\Array{ccc}{ \frac{1}{\sqrt{1+l_1^2}} & 0 & -\frac{l_1}{\sqrt{1+l_1^2}} \\	
		-\frac{l_1l_2}{\sqrt{1+l_1^2}\sqrt{1+l_1^2+l_2^2}} & \frac{\sqrt{1+l_1^2}}{\sqrt{1+l_1^2+l_2^2}} & -\frac{l_2}{\sqrt{1+l_1^2}\sqrt{1+l_1^2+l_2^2}} \\
		\frac{l_1}{\sqrt{1+l_1^2+l_2^2}} & \frac{l_2}{\sqrt{1+l_1^2+l_2^2}} & \frac{1}{\sqrt{1+l_1^2+l_2^2}}}.
\label{eq.Sdef}
\end{eqnarray}
Taking into account the $SO(2)^q$ symmetry, a general diagonalization can be written as
\begin{eqnarray}
	\Big[S^qR_n(\theta)\Big]M_N^q\Big[S^qR_n(\theta)\Big]^T=\textrm{diag}(0,0,1).
\label{eq.SRMSR}
\end{eqnarray}
Notably
\begin{eqnarray}
S^qR_n(\theta)(S^q)^T=R_3(\theta)=\Array{ccc}{c_\theta & s_\theta & 0 \\ 
		-s_\theta & c_\theta & 0 \\
		0 & 0 & 1},
\label{eq.SRSR3}
\end{eqnarray} 
it implies that the $SO(2)^q$ rotation angle $\theta^q$ selects a superposition basis within the degenerate mass space of the first two families in the hierarchy limit.

\subsection{Structure of the CKM Matrix}
Combining Eqs.  (\ref{eq.factorizedM}), (\ref{eq.SRMSR}), and (\ref{eq.SRSR3}),  the CKM matrix takes the form
\begin{eqnarray}
V_{CKM}= R_3(\theta^u)S^u\Array{ccc}{e^{i\lambda_1} && \\ &e^{i\lambda_2}& \\ &&1}(S^d)^TR_3^T(\theta^d)
\label{eq.Vckmstru}
\end{eqnarray}
where $\lambda_i\equiv \eta_i^d-\eta_i^u$. 
Not all Yukawa phases $\eta^q_i$ in Eq. (\ref{eq.KLdef}) are physical; Only the difference $\eta_i^d-\eta_i^u$ appear in observable quantities.
For fixed patterns $l_1,l_2$ for up- and down-type quarks, Eq. (\ref{eq.Vckmstru}) contains just four parameters, $\theta^u,\theta^d,\lambda_1$, and $\lambda_2$. These four parameters correspond precisely to the three mixing angles and one CP-violating phase, with no redundancy. 
This indicates that the charged weak current breaks the $SO(2)^u\times SO(2)^d$ family symmetry, meaning that $\theta^u$ and $\theta^d$ are selected by the CKM mixing. Otherwise, the quark mixing would exhibit approximate $SO(2)^u\times SO(2)^d$ symmetry.

\subsection{Unification Through Sub-Unitarity}
A successful quark flavor structure must simultaneously incorporate the factorized mass matrix and the sub-unitarity of CKM mixing. 
In Eq. (\ref{eq.Vckmstru}), the sub-unitarity condition in the hierarchy limit imposes two requirements
\begin{eqnarray}
\lambda_1=\lambda_2=0,~~~~S^u(S^d)^T=1.
\label{eq.subuCons}
\end{eqnarray}
The first requirement indicates that the Yukawa phases arise from the hierarchy corrections, consistent with the vanishing of  CP violation in the two-family limit. It also implies that  $\lambda_{1}$ and $\lambda_2$ must be small in any successful phenomenological fit. 

The second requirement points toward a unified mass pattern. Although $S^d$ and $S^d$ are generally derived from mass matrices 
$M^u$ and $M^d$ with potentially different choices of the parameters $l_1$ and $l_2$, 
Eq. (\ref{eq.subuCons}) demands that $M_N^u$ and $M_N^d$ share the same $l_1,l_2$ to achieve $S^u=S^d$. 
Hence, the up- and down-type quark mass matrices are homogeneous, governed by identical pattern matrices.
Under these two conditions, the CKM matrix in the hierarchy limit reduces to
\begin{eqnarray}
\lim_{h_{23}^{u,d}\rightarrow0}V_{CKM}= R_3(\theta^u)R_3^T(\theta^d)=\Array{ccc}{\cos(\theta^u-\theta^d) & \sin(\theta^u-\theta^d) & \\
 -\sin(\theta^u-\theta^d) & \cos(\theta^u-\theta^d) & \\
 && 1}
.
\end{eqnarray}
Now, the sub-unitary $V_{CKM}$ is parameterized by a single rotation angle 
$\theta_{12}=\theta^u-\theta^d$, 
 which exactly matches the two-family mixing in Eq. (\ref{eq.2Fmixing0}). 
This result provides a powerful self-consistent test: in any successful fit,  $\theta^u$ and $\theta^d$ should lie approximately along a straight line with slope $\arctan (\theta_{12})$ crossing the origin of coordinates in the $\theta^u-\theta^d$ plane.

\subsection{The Flat Matrix as the Natural Pattern}
The unification condition $S^u=S^d$ requires identical $l_1,l_2$ for up- and down-type quarks, but does not specify their values. Suppose two pattern matrices $M_N^q$ and ${M_N^q}^\prime$, governed by $(l_1,l_2)$ and $(l_1^\prime,l_2^\prime)$, are digaonalized by $S^q$ and ${S^q}^\prime$, respectively. An orthogonal transformation relates them 
\begin{eqnarray}
M_N^q=[(S^q)^TS'^q] M_N'^q[(S^q)^TS'^q]^T.
\end{eqnarray}
Thus, two mass patterns produce equivalent mass eigenvalues. 
However, they yield different CKM matrices through 
Eq. (\ref{eq.Vckmstru}).
Phenomenological fitting can determine the allowed parameter regions for $l_1$ and $l_2$. 
A recent study \cite{QCaoNPB2025} examined the allowed ranges of these parameters and concluded that the constraints from CKM mixing impose almost no restrictions on $l_1$ and $l_2$ at the $2\sigma$ C.L. 
This scenario raises a puzzle question: why would nature select a specific set of 
$l_1,l_2$?
The values of $l_1$ and $l_2$ can not be arbitrary theoretically. Rather, they reflect a more fundamental principle. Since quark masses in the Standard Model originate from Yukawa interactions, a clear and organized structure for the Yukawa couplings is essential in the final flavor framework.
While numerous combinations of $l_1$ and $l_2$ are consistent with phenomenological observations, only an elegant assignment is likely to encapsulate the underlying Yukawa interaction. This assignment is given by $l_1=l_2=1$. 
In this case, $M_N^q$ takes the remarkably simple form of a flat matrix with all entries equal to 1, which we denote by ${M_F^q}$
\begin{eqnarray}
	M_F^q=\frac{1}{3}\Array{ccc}{1&1&1 \\1&1&1 \\1&1&1}.
\end{eqnarray}
Any other assignments would require an additional explanation for why nature chose those particular values. The flat pattern, by contrast, embodies maximal simplicity and harmony: Yukawa interactions between different families become identical. In the Yukawa basis, the Yukawa interaction in the hierarchy limit is expressed by
\begin{eqnarray}
\mathcal{L}_Y=-y^u\sum_{i,j=1,2,3}\bar{q}^{(Y)}_{L,i}\tilde{H}u^{(Y)}_{R,j}
		-y^d\sum_{i,j=1,2,3}\bar{q}^{(Y)}_{L,i}{H}d^{(Y)}_{R,j}.
\end{eqnarray}
In the rest of this paper, we concentrate on this flat pattern, considering it the most natural candidate for a unified flavor structure.

\section{Hierarchy Corrections}
\label{sec.HighOrderCorrection}
Building on the mass pattern and CKM structure in the hierarchy limit, we now turn to the masses of the light quarks and the precise corrections to the CKM matrix. 

\subsection{Corrections to Mass Pattern}
The mass pattern matrix $M_N^q$ yields the normalized spectrum $(0,0,1)$ in the hierarchy limit. Any deviation from the conditions in Eq. (\ref{eq.MNrelation1}) and (\ref{eq.MNrelation2}) will shift this spectrum. We focus on corrections to the flat pattern $M_F^q$ and denote the corrected pattern matrix as $M_\delta^q$. Before parameterizing $M_\delta^q$, the following two considerations guide us
\begin{itemize}
	\item[1.] $M_\delta^q$ must remain real symmetric to preserve orthogonal diagonalization. Non-symmetric correction would break the hermiticity established by fixing $U_R^q=U_L^q$ in Sec. \ref{sec.generallFramework}.
	\item[2.] Corrections to diagonal elements of $M_\delta^q$ can be transformed into non-diagonal elements by an orthogonal rotation $\tilde{R}$:
		\begin{eqnarray}
			\tilde{R}\Array{ccc}{ 1+\Delta_1 & 1 & 1 \\
			 1 & 1+\Delta_2 & 1 \\
			 1 & 1 & 1+\Delta_3}\tilde{R}^T=\Array{ccc}{ 1 & 1 +\Delta'_1 & 1+\Delta'_2 \\
			 1 +\Delta'_1& 1 & 1+\Delta'_3 \\
			 1 +\Delta'_2& 1+\Delta'_3 & 1}
		\end{eqnarray}
	where $\Delta_i$ and $\Delta'_i$ are real corrections.
\end{itemize}

Thus, the general corrected mass matrix can be parameterized as an off-diagonal real symmetric matrix
\begin{eqnarray}
M_\delta^q=\frac{1}{3}\Array{ccc}{ 1 & 1 +\delta_{12}^q & 1+\delta_{13}^q \\
			 1 +\delta_{12}^q& 1 & 1+\delta_{23}^q \\
			 1 +\delta_{13}^q& 1+\delta_{23}^q & 1}.
\label{eq.MdeltaDef}
\end{eqnarray}

\subsection{$h^1$ and $h^2$ Order Corrections}
The quark masses follow the hierarchical pattern
\begin{eqnarray}
	\frac{m_1^q}{\sum_im_i^q}&=&h_{12}^qh_{23}^q\sim \mathcal{O}(h^2),
	\\
	\frac{m_2^q}{\sum_im_i^q}&=&h_{23}^q\sim \mathcal{O}(h^1),
	\\
	\frac{m_3^q}{\sum_im_i^q}&=&1-h_{23}^q-h_{12}^qh_{23}^q\sim \mathcal{O}(1).
\end{eqnarray}
The lighter quark masses arise from shifting the eigenvalues: $(0,0,1)\rightarrow(0, h_{23}^q,1-h_{23}^q)$ at $\mathcal{O}(h^1)$.
Solving from the corrections $\delta_{ij}^q$ order by order \cite{ZhangEPJC2025}, we obtain $\delta_{ij}^q$ at $\mathcal{O}(h^1)$
\begin{eqnarray}
	\delta_{12}^q&=&\Big(-\frac{3}{4}\cos(2\theta^q)-\frac{9}{4\sqrt{3}}\sin(2\theta^q)-\frac{3}{2}\Big)h_{23}^q,
	\\
	\delta_{23}^q&=&\Big(-\frac{3}{4}\cos(2\theta^q)+\frac{9}{4\sqrt{3}}\sin(2\theta^q)-\frac{3}{2}\Big)h_{23}^q,
	\\
	\delta_{13}^q&=&\Big(\frac{2}{3}\cos(2\theta^q)-\frac{3}{2}\Big)h_{23}^q.	
\end{eqnarray}
Here, $\theta^q$ is the $SO(2)^q$ rotation angles. It indicates that the $SO(2)^q$ family symmetry remains approximately. This symmetry can be made explicit by expressing $M_\delta^q$ as a function of $\theta^q$:
\begin{eqnarray}
M_\delta^q(\theta^q)=R_\delta^TM_\delta^q(0)R_\delta(\theta^q)
\label{eq.RMRdelta}
\end{eqnarray}
where 
$R_\delta(\theta)$ is $SO(2)^q$ rotation along the corrected axial in the direction $(1,1-\frac{9}{4}h_{23}^q,1)$ and $M_\delta^q(0)$ is the corrected mass matrix at $\theta^q=0$.

After some tedious calculation, the orthogonal transformation $S_\delta^q$ that diagonalizes $M_\delta^q(0)$ is
\begin{eqnarray}
	S_\delta^q=\Array{ccc}{\frac{1}{\sqrt{2}} & 0 & -\frac{1}{\sqrt{2}} \\
		-\frac{1}{\sqrt{6}} & \frac{\sqrt{2}}{\sqrt{3}} & -\frac{1}{\sqrt{6}} \\
		\frac{1}{\sqrt{3}} & \frac{1}{\sqrt{3}} & \frac{1}{\sqrt{3}} }
		+\frac{h_{23}^q}{4\sqrt{3}}\Array{ccc}{ 0 & 0 & 0 \\ \sqrt{2} & \sqrt{2} & \sqrt{2} \\ 1 & -2  & 1}+\mathcal{O}(h^2).
\end{eqnarray}
It satifies 
\begin{eqnarray}
	S_\delta^qM_\delta^q(0){S_\delta^q}^T=\textrm{diag}(0,h_{23}^q,1-h_{23}^q).
\label{eq.SMSdelta}
\end{eqnarray}

To accommodate the lightest quark mass, $\mathcal{O}(h^2)$ corrections must be included.
Following the same approach, $\delta_{ij}^q$ and $S_\delta^q$ can be extended to $\mathcal{O}(h^2)$ order, incorporating terms of $(h_{23}^q)^2$ and $h_{12}^qh_{23}^q$. The $SO(2)^q$  symmetry remains valid up to $\mathcal{O}(h^2)$. Detailed formulas are listed as follows
\begin{eqnarray}
	\delta_{12}^q&=&\Big[-\frac{3}{4}\cos(2\theta^q)-\frac{3\sqrt{3}}{4}\sin(2\theta^q)-\frac{3}{2}\Big]h_{23}^q-3h_{12}^qh_{23}^q
	-\frac{9}{32}\Big[2\cos(2\theta^q)+1\Big]^2(h_{23}^q)^2+\mathcal{O}(h^3),
	\label{eq.deltah21}\\
	\delta_{23}^q&=&\Big[-\frac{3}{4}\cos(2\theta^q)+\frac{3\sqrt{3}}{4}\sin(2\theta^q)-\frac{3}{2}\Big]h_{23}^q-3h_{12}^qh_{23}^q
	-\frac{9}{32}\Big[2\cos(2\theta^q)+1\Big]^2(h_{23}^q)^2+\mathcal{O}(h^3),
	\label{eq.deltah22}\\
	\delta_{13}^q&=&\Big[\frac{2}{3}\cos(2\theta^q)-\frac{3}{2}\Big]h_{23}^q-3h_{12}^qh_{23}^q+\mathcal{O}(h^3).	
	\label{eq.deltah23}
\end{eqnarray}

\subsection{Corrections to Mixing}
Using Eqs. (\ref{eq.deltah21}), (\ref{eq.deltah22}) and  (\ref{eq.deltah23}), the $M_\delta^q(\theta^q)$ is generally diagonalized by $S_\delta^qR_\delta(\theta^q)$ as
\begin{eqnarray}
\Big[S_\delta^qR_\delta(\theta^q)\Big]M_\delta^q(\theta^q)\Big[S_\delta^qR_\delta(\theta^q)\Big]^T=\textrm{diag}(h_{12}^qh_{23}^q,h_{23}^q,1-h_{23}^q)+\mathcal{O}(h^3).
\end{eqnarray}

Thus, the CKM mixing matrix up to $\mathcal{O}(h^2)$ can be written as:
\begin{eqnarray}
V_{CKM}=\Big[S_\delta^uR_\delta(\theta^u)\Big]\textrm{diag}(e^{i\lambda_1},e^{i\lambda_2},1)\Big[S_\delta^dR_\delta(\theta^d)\Big]^T.
\label{eq.CKMdelta}
\end{eqnarray}
As a self-consistent check, in the limit of $h_{23}^{q}\rightarrow0$, we have
\begin{eqnarray}
\lim_{h_{23}^{q}\rightarrow 0}R_\delta(\theta^q)&=&R_N(\theta^q)
\\
\lim_{h_{23}^{q}\rightarrow 0}S_\delta^q&=&S_0^q
\end{eqnarray}
where $R_N(\theta)$ is a $SO(2)$ rotation along the axial in the direction $(1,1,1)$.
Thus, Eq. (\ref{eq.CKMdelta}) correctly recovers the hierarchy limit expression in Eq. (\ref{eq.Vckmstru}).

\section{Phenomenological Fits}
\label{sec.fit}
Quark masses and CKM mixing represent the two complementary aspects of flavor phenomenology. Any candidate flavor structure must successfully reproduce both. In previous sections, we established the flat pattern and CKM structure based on quark mass hierarchy. Here, we perform detailed fits to experimental data.

Using quark mass data in Tab. \ref{tab.quarkmass},  the hierarchy parameters are
\begin{eqnarray}
&&
h_{23}^u=7.377\times 10^{-3},
~~~
h_{12}^u=1.697\times 10^{-3},
\label{eq.h12h23data1}
\\
&&
h_{23}^d=2.235\times 10^{-2},
~~~
h_{12}^d=5.027\times 10^{-2}.
\label{eq.h12h23data2}
\end{eqnarray}
For the corrected mass matrix $M_\delta^q(\theta)$ given in Eq. (\ref{eq.MdeltaDef}), its eigenvalues can be expressed as functions of 
$\theta$.  Through numerical calculations, it is verified that the results possess an approximate $SO(2)^q$ symmetry.

The key test of our flavor structure is the CKM fit.
Using the $\mathcal{O}(h^2)$ corrections, we input the hierarchy parameters $h_{ij}^q$  with their experimental values into 
$V_{CKM}$  in Eq. (\ref{eq.CKMdelta}).
The fitting task is to find a set of parameters $(\theta^u,\theta^d,\lambda_1,\lambda_2)$ that yields CKM mixing angles and CP phase consistent with experimental data.
To verify the sub-unitarity implication, a successful fit additionally requires the approximate linear relation between $\theta^u$ and $\theta^d$, as well as small Yukawa phases $\lambda_1$ and $\lambda_2$.

Scanning all parameter space of $(\theta^u,\theta^d,\lambda_1,\lambda_2)$, all allowed parameters are recorded when the CKM mixing data falls within the range of $2\sigma$ C.L. 
\begin{table}[htp]
\caption{Fit Result}
\begin{center}
\begin{tabular}{c|c|c}
\hline
\hline
para. & CKM exp. & Fit  
\\
\hline
 $\begin{array}{c}
 	\theta^u=4.681 \\
	\theta^d=4.502 \\
	 \lambda_1=-0.1028 \\
	 \lambda_2=-0.04696\end{array}$
&
$\begin{array}{c}
	s_{12}=0.22501\pm0.00068 \\ 
	s_{23}=0.04183^{+0.00079}_{-0.00069} \\
	 s_{13}=0.003732^{+0.000090}_{-0.000085} \\ 
	 \delta_{CP}=1.147\pm0.026\end{array}$
 & $\begin{array}{c}
 	s_{12}=0.2247 \\
 	s_{23}=0.04228 \\ 
	s_{13}=0.003654 \\
	\delta_{CP}=1.13807
	\end{array}$
\\
\hline
\hline
\end{tabular}
\end{center}
\label{tab.fitresult}
\end{table}
The fit results verify the sub-unitarity prediction: (1)
 In the plane of $\theta^u-\theta^d$,  fit points cluster around the line $\theta^u-\theta^d=\theta_{12}$, corresponding to the two-family rotation angle; 
(2) In the plane of $\lambda_1-\lambda_2$, points concentrate near the origin, confirming that Yukawa phases are small, consistent with CP violation arising from hierarchy corrections.
These fits demonstrate that the flat pattern, combined with hierarchy corrections, successfully reproduces all quark flavor observables while maintaining the sub-unitarity condition as a fundamental organizing principle.


\section{Summary}
\label{sec.sum}
We have derived two fundamental implications about quark flavor structure from the hierarchy of quark masses: (1) in the hierarchy limit, quark mass matrices factorize into a diagonal phase matrix and a pattern matrix controlled by two real parameters; (2) the large mass gap between the third and first two families implies that quark mixing approximately reduces to a unitary $2\times2$ rotation below the third-family mass scale. 
Combining these implications yields a successful quark-flavor framework that establishes a clear relation between quark masses and mixings, explains the origin of $\theta_{13}$ and $\theta_{23}$ from hierarchy corrections, and passes precision phenomenological tests. These features make the flat quark flavor structure a strong candidate to replace the ambiguous Yukawa couplings of the Standard Model.

The factorized mass matrix can then be extended to the lepton sector for three generations of normal ordering Dirac neutrinos satisfying $h_{23}^\nu\ll 1$. However, the sub-unitarity mechanism does not directly apply to leptons due to the smallness of neutrino masses. This explains why the lepton PMNS mixing matrix exhibits two large angles, in contrast to the quark sector, where the small mixing angles $\theta_{13}$ and $\theta_{23}$ arise from hierarchy corrections. Whether this difference between quarks and leptons can be explained by a common underlying flavor structure remains an open question for future investigation.

\begin{acknowledgments}
This work is supported by Shaanxi Natural Science Foundation 2022JM-052.
\end{acknowledgments}

\end{document}